\documentclass[11pt]{article}
\usepackage{amsfonts,epsfig,latexsym}

\newcommand{\R}{{\mathbb{R}}}
\newcommand{\Z}{{\mathbb{Z}}}

\newcommand{\C}{{\mathbb{C}}}
\newcommand{\beq}{\begin{equation}}
\newcommand{\eeq}{\end{equation}}
\newcommand{\ra}{\rightarrow}

\newcommand{\cd}{\partial}
\newcommand{\wt}{\widetilde}

\newcommand{\nvec}{{\bf n}}
\newcommand{\evec}{{\bf e}}
\newcommand{\uvec}{{\bf u}}
\newcommand{\xvec}{{\bf x}}

\newcommand{\epsvec}{\mbox{\boldmath{$\varepsilon$}}}
\newcommand{\cs}{\cos\theta}
\newcommand{\sn}{\sin\theta}
\newcommand{\fl}{{\cal F}\ell}
\newcommand{\spec}{{\rm spec}\, }
\newcommand{\spn}{{\rm span}\, }
\newcommand{\diag}{{\rm diag}\, }
\newcommand{\re}{{\rm Re}\, }
\newcommand{\rot}{{\cal R}}

\newtheorem{thm}{Theorem}

\newcommand{\be}{\begin{equation}}
\newcommand{\ee}{\end{equation}}
\newcommand{\bea}{\begin{eqnarray}}
\newcommand{\eea}{\end{eqnarray}}
\newcommand{\bean}{\begin{eqnarray*}}
\newcommand{\eean}{\end{eqnarray*}}

\newcommand{\news}{\setcounter{equation}{0}\hskip 0.5cm}
\newcommand{\bn}{{\bf n}}
\newcommand{\vac}{{\bf e}_3}

\begin{document}

\title{Discrete Breathers in Anisotropic Ferromagnetic Spin Chains}
\author{J.M. Speight\thanks{Email : J.M.Speight@leeds.ac.uk} \\ 
Department of Pure Mathematics, 
University of Leeds, \\ Leeds LS2 9JT, U.K. \\
\\
\\ P.M. Sutcliffe\thanks{Email : P.M.Sutcliffe@ukc.ac.uk} \\ 
Institute of Mathematics, University of Kent at Canterbury,\\
Canterbury, CT2 7NF, U.K.}
\date{}
\maketitle

\begin{abstract}
\noindent 
We prove the existence of discrete breathers (time-periodic, spatially localized solutions)
in weakly coupled ferromagnetic spin chains with easy-axis anisotropy. Using numerical
methods we then investigate the continuation of discrete breather solutions as the intersite
coupling is increased. We find a band of frequencies for which the 1-site
breather continues all the way to the soliton solution in the continuum. There is a second
band, which abuts the first, in which the 1-site breather does not continue to the 
soliton solution, but a certain multi-site breather does. This banded structure continues,
so that in each band there is a particular multi-site breather which continues to the soliton
solution. A detailed analysis is presented, including an exposition of how the bifurcation
pattern changes as a band is crossed. The linear stability of breathers is analyzed. It is
proved that 1-site breathers are stable at small coupling, provided a non-resonance condition
holds, and an extensive numerical stability analysis of 1-site and multisite breathers is
performed. The results show alternating bands of stability and instability as the coupling 
increases.

\end{abstract}
\newpage
\section{Introduction}\news

Discrete breathers are time periodic, spatially localized solutions in
networks of coupled oscillators (including rotors and spins). They arise in
a variety of very general systems \cite{aubmac,SM} due to the interplay between nonlinear and
discrete effects, and therefore there is an enormous potential for their application
in many areas of physics, particularly condensed matter and biophysics, where physical
systems are often inherently discrete. Generally one finds that as the system is moved 
closer to its continuum limit, by increasing a coupling constant in the theory, 
there comes a point at which 
a discrete breather solution no longer exists. This is to be expected since an 
increase in the coupling constant results in an expansion of the phonon frequency band
and eventually this band captures a harmonic of the breather frequency. In fact, it is
often the case that the discrete breather cannot be continued even up to the point at
which the above resonance argument applies, and this is not well understood.

In this paper we study the classical equations of motion for a ferromagnetic
spin chain with an easy-axis anisotropy. Some numerical studies of discrete breathers
(which are known as intrinsic localized modes in the condensed matter physics
literature) have been performed in the cases of  easy-plane ferromagnets \cite{WMB} and
 easy axis antiferromagnets \cite{laisie}. However,
both the perspective and the results of the current investigation are quite different.
We prove the existence of discrete breathers in the weakly coupled case by starting
from the anti-continuum limit (zero coupling constant) and applying an implicit
function theorem argument to the trivial 1-site breather. We then apply numerical
methods to investigate the continuation of this solution as the coupling constant 
is increased, with the novel result that it continues for all values of the coupling.
This is explained by an examination of the spin waves.
Depending upon the breather frequency, the continuation results in either the soliton
solution of the continuum model or a trivial static solution in which all spins are
either aligned or anti-aligned with the vacuum. Similar results are found to apply
to multi-site breathers, which fit into an intricate pattern, and lead us to conjecture
that the continuum soliton solution (of any allowed frequency) can always be obtained 
from the continuation of a particular form of multi-site breather. Evidence in support
of this conjecture is presented. 
The numerical results are in some ways reminiscent of
those found \cite{eil1,dnls} for the discrete nonlinear Schr\"odinger (DNLS) system, but have
not previously been seen in spin chains.

We then go on to perform a linear stability analysis of the breathers within the Krein theory
framework pioneered by Aubry \cite{aubstab}. We prove that nonresonant continued 
1-site breathers must be stable for sufficiently weak coupling, and numerically investigate
the stability of both 1-site and multisite breathers. The results suggest that those families
which tend to the continuum soliton experience regular repeating bands of instability of
diminishing strength as the coupling increases. 

During the refereeing process for this paper, a very interesting 
paper by Flach, Zolotaryuk and Fleurov
on breathers in very similar systems appeared \cite{flazol}. The differences between their
results and ours are, briefly, as follows. They consider both easy-axis and easy-plane
anisotropy, but in the case where the exchange interaction is anisotropic also (our exchange
interaction maintains isotropy). Breather existence is proved by continuing from a limit in 
which two of the exchange integrals vanish, but the third remains nonzero. Since the 
continuation is for small values of the continuation parameters (the two small exchange 
integrals) their existence result applies to strongly anisotropic exchange interaction, as
opposed to ours, which applies to isotropic exchange. More importantly, in this paper we
exploit the isotropic exchange to reduce the breather equations to a purely algebraic system,
which greatly simplifies the analysis in comparison with theirs.
Consequently our existence result requires none of the nonresonance hypotheses of Flach
et al's. Also we are able to build spatial localization of breathers directly into our
analysis, whereas Flach et al do not consider this issue at all (although standard results
of MacKay on exponential breather localization should apply to their system just as they do to
ours - see section \ref{sec-analytic}). 
 Similarly, we exploit the exchange isotropy to simplify the linear stability analysis,
again reducing it to a purely algebraic problem, a reduction not possible for the systems
analyzed in \cite{flazol}. We are able, therefore, to perform a very detailed numerical 
existence and
stability analysis at minimal computational cost. To summarize, Flach et al prove a result
of rather more general applicability and physical relevance than ours, 
though their context does not, strictly
speaking, include ours. We consider a somewhat more idealized system, but obtain correspondingly
stronger results.

\section{Anisotropic ferromagnetic spin chains}\news
\label{defns}

The classical formulation of a spin chain involves a three-component
unit vector ${\bf n}_i$, giving the spin at each lattice site $i\in\Z.$ 
The type of spin chain is defined by its Hamiltonian, which we take to be
\be
H=\sum_i\{\alpha(1-{\bf n}_i\cdot{\bf n}_{i+1})+\frac{A}{2}[1-({\bf n}_i\cdot {\bf e}_3)^2]\}.
\label{ham}\ee
Here $\alpha\ge 0$ is the coupling constant (exchange integral), which is positive
in the case of a ferromagnet, and $A>0$ is the anisotropy constant, which is also
positive since we wish to consider an easy-axis anisotropy,
 with ${\bf e}_3=(0,0,1)$ being the easy-axis. In this case the 
minimum of the Hamiltonian is zero, obtained by the vacuum configuration
$\bn_i=\pm \vac.$ In the following we shall choose $+\vac$ to be the vacuum
configuration and refer to spins which take the values $+\vac$ and $-\vac$
as being spin up and down respectively. Furthermore, by applying a scaling symmetry
we can, without loss of generality, set $A=1.$

The equation of motion is obtained from the Hamiltonian as
\be
\dot\bn_i=-\bn_i\times\frac{\partial H}{\partial \bn_i}
\label{eom1}
\ee
where a dot denotes differentiation with respect to time.
Using (\ref{ham}) we obtain
\be
\dot\bn_i=\alpha\bn_i\times(\bn_{i+1}+\bn_{i-1})+(\bn_i\cdot\vac)(\bn_i\times\vac).
\label{eom2}
\ee
Discrete breather solutions have the form
\be
\bn_i(t)=(\sin\theta_i\cos\omega t,-\sin\theta_i\sin\omega t,\cos\theta_i)
\label{ansatz}
\ee
where $\omega$ is the frequency. Substituting this ansatz into the equation of
motion (\ref{eom2}) yields the following 
nonlinear second order difference equation for the angles $\theta_i$,
\be
\alpha\{\cos\theta_i(\sin\theta_{i+1}+\sin\theta_{i-1})
-\sin\theta_i(\cos\theta_{i+1}+\cos\theta_{i-1})\}
=\sin\theta_i(\cos\theta_i-\omega).
\label{angeom}
\ee
Such a periodic solution may properly be called a discrete breather if it is
spatially localized, that is, $\lim_{i\ra\pm\infty}\theta_i=0$.

In section \ref{sec-analytic} we shall prove the existence of discrete
breathers for $\alpha$ sufficiently small and
in section \ref{sec-numeric} we shall study them numerically for a range
of values of $\alpha.$

Note that for all $\alpha$ there is a set of static solutions in which
each spin can independently be chosen to point either up or down, since
in each case we have that $\sin\theta_i=0$ for all $i,$ which clearly solves equation
(\ref{angeom}). These solutions will play an important role later in our discussion.

Finally, in this section, we address the continuum limit $\alpha\rightarrow\infty.$
Write $\alpha=1/h^2$ and regard $\bn_i$ as the value of a continuous function 
$\bn(x)$ sampled at the lattice points $x=ih.$ Taking the continuum limit
$h\rightarrow 0$, the Hamiltonian (\ref{ham}) becomes
\be
H=\int\{ \frac{1}{2}\bn'\cdot\bn'+ \frac{1}{2}[1-({\bf n}_i\cdot {\bf e}_3)^2]\} \ dx
\label{hamc}\ee
and the equation of motion is
\be
\dot\bn=\bn\times\bn''+(\bn\cdot\vac)(\bn\times\vac).
\label{eom2c}
\ee
where prime denotes differentiation with respect to $x.$

This partial differential equation has time periodic, exponentially localized solutions
known as magnetic solitons \cite{KIK}. At this point it is perhaps worth pointing out that
there is an unfortunate difference in nomenclature, in that in the discrete case 
time periodic localized solutions are known as breathers, whereas in the continuum model
they are termed solitons. We shall continue to use the term soliton when referring to the
continuum limit, but the reader should be aware that it has exactly the same time
dependence as the discrete breather and moreover, as we shall see later, the soliton
can be obtained from the discrete breather in the continuum limit.

Using the same form for the time dependence as in the discrete case (\ref{ansatz}) 
i.e.
\be
\bn(x,t)=(\sin\theta(x)\cos\omega t,-\sin\theta(x)\sin\omega t,\cos\theta(x))
\label{ansatzc}
\ee
equation (\ref{eom2c}) yields the second order ordinary differential equation
\be
\theta''=(\cos\theta-\omega)\sin\theta
\label{angeomc}
\ee
which, of course, is also obtained from the continuum limit of (\ref{angeom}).
The boundary conditions for a soliton solution, located at the origin,
are $\theta'(0)=\theta(\infty)=0.$ Equation (\ref{angeomc}) can be integrated
explicitly and the solution satisfying the correct boundary conditions is given
by
\be
\theta(x)=\cos^{-1}\{
\frac{2\omega}{1-(1-\omega)\tanh^2(x\sqrt{1-\omega})}-1\}.
\label{soliton}\ee
Since we require the soliton to be a smooth exponentially localized solution,
this formula shows that the frequency must be restricted to the range $\omega\in(0,1).$
There is a second type of soliton solution \cite{KIK} for which $\omega<0$,
but this has a different structure from (\ref{soliton}); in particular
$\bn=-\vac$ at the centre of the soliton for all $\omega.$ This negative frequency
soliton will not arise in our discussion, so in this paper when we refer to
a soliton we shall mean the solution (\ref{soliton}).

\section{Analytic results on discrete breathers}\news
\label{sec-analytic}
Following 
the ``homoclinic orbit'' approach of Flach \cite{fla}, one would like to
prove directly the existence of solutions of (\ref{angeom}) with the correct
boundary behaviour, using techniques of dynamical systems theory. 
Unfortunately, equation (\ref{angeom}) does not determine a well-defined 
homeomorphism of the torus, $(\theta_{i-1},\theta_i)\mapsto (\theta_i,
\theta_{i+1})$, so the direct approach is not convenient here.

However, we may still prove the existence of breathers in this system by
continuation of 1-site breathers from the decoupled limit ($\alpha=0$),
in the manner of Aubry and MacKay's work on oscillator networks
\cite{aubmac}. 
Here we have an algebraic system rather than an infinite  system of ODEs so the details
are considerably less technical (compare, for example, with
\cite{flazol}), and will be treated with corresponding brevity.
The idea is that when $\alpha=0$, $\omega\in(-1,1)$, 
(\ref{angeom}) supports the almost trivial solution
\beq
\label{m4}
\theta^b_i=\left\{\begin{array}{cl}
0 & i\neq 0 \\
\cos^{-1}\omega & i=0
\end{array}\right.
\eeq
in which every spin remains pointing up except one (whose
location we have chosen to be $i=0$) which precesses with frequency 
$\omega$ around
a circle of fixed latitude. Note that there is no reason to assume that
$\omega>0$ as we had to in the continuum system. Keeping $\omega$ fixed,
the existence of breathers for all $\alpha$ sufficiently small will follow
from an implicit function theorem argument.

To be precise, let $F:\ell_2\oplus\R\ra\ell_2$ such that
\beq
\label{m5}
[F(\theta,\alpha)]_i=(\cs_i-\omega)\sn_i-\alpha[\cs_i(\sn_{i+1}
+\sn_{i-1})-\sn_i(\cs_{i+1}+\cs_{i-1})],
\eeq
where $\ell_2$ is the Banach space of sequences $\theta:\Z\ra\R$ with finite
norm
\beq
\label{m6}
||\theta||_{\ell_2}=\left[\sum_i\theta_i^2\right]^{\frac{1}{2}}.
\eeq
Note that $\theta\in\ell_2$ is a solution of (\ref{angeom}) at coupling $\alpha$
if and only if $F(\theta,\alpha)=0$. Note also that every $\theta\in\ell_2$
converges to $0$ as $i\ra\pm\infty$, so such a zero of $F(\cdot,\alpha)$ is
a discrete breather. For all $\omega\in(-1,1)$, $F(\theta^b,0)=0$, where
$\theta^b$ is the 1-site breather defined in (\ref{m4}), and the partial
differential of $F$ with respect to the $\ell_2$ factor at $(\theta,\alpha)=
(\theta^b,0)$ is easily seen to be an isomorphism
$DF_{(\theta^b,0)}:\ell_2\ra\ell_2$, that is, a bijection with bounded
inverse. In fact this linear map is diagonal:
\beq
\label{m7}
[DF_{(\theta^b,0)}\, \delta\theta]_i=\left\{
\begin{array}{rl}
(1-\omega)\delta\theta_i & i\neq 0 \\
-(1-\omega^2)\delta\theta_0 & i=0.
\end{array}
\right.
\eeq
Hence we may apply the implicit function theorem \cite{ift} to obtain the
following
\begin{thm}\label{exist} 
For all $\omega\in(-1,1)$ there exist $\epsilon>0$ and a
$C^1$ map $[0,\epsilon)\ra\ell_2$, denoted $\alpha\mapsto \theta^\alpha$,
such that $\theta^\alpha$ is a frequency $\omega$, coupling $\alpha$ solution
of (\ref{angeom}) and $\theta^0=\theta^b$. The map $\alpha\mapsto\theta^\alpha$
is unique provided $\epsilon$ is chosen sufficiently small.
\end{thm}

As noted above, since the continuation occurs in $\ell_2$, it has weak
spatial localization built in. This can be improved to exponential
spatial localization (that is, there exist $C,\lambda>1$ such that
$|\theta_i|<C\lambda^{-|i|}$) by applying some standard results of
Baesens and MacKay. The idea is to differentiate the continuation equation
$F(\theta^\alpha,\alpha)=0$ with respect to $\alpha$ to obtain an ODE
on $\ell_2$, namely,
\beq
\label{m8}
\frac{d\theta^\alpha}{d\alpha}=[DF_{(\theta^\alpha,\alpha)}]^{-1}
\Delta(\theta^\alpha),
\eeq
where $\Delta:\ell_2\ra\ell_2$ such that
\beq
\label{m9}
[\Delta(\theta)]_i=\cs_i(\sn_{i+1}
+\sn_{i-1})-\sn_i(\cs_{i+1}+\cs_{i-1}).
\eeq
The theorem above may be interpreted as asserting the local existence and
uniqueness of a solution to the initial value problem $\theta^0=\theta^b$
for (\ref{m8}). Now $\Delta(\theta^\alpha)$ is clearly exponentially 
localized provided $\theta^\alpha$ is, as is 
$[DF_{(\theta^\alpha,\alpha)}]^{-1}
\Delta(\theta^\alpha)$ by theorem 3 of \cite{baemac}. Thus one obtains
an exponential decay estimate for $d\theta^\alpha/d\alpha$ at $\alpha_0$
given a decay estimate for $\theta^{\alpha_0}$. This may be integrated to
show that the estimate for $\theta^{\alpha_0}$, if chosen correctly,
 remains valid in a finite interval containing $\alpha_0$
(see the proof of theorem 2 of \cite{baemac}, for example). 
It remains then to
note that the initial datum $\theta^b$ trivially satisfies {\em all}
exponential decay criteria (i.e.\ one can choose any $\lambda>1$), 
so exponential localization persists for $\alpha$ sufficiently small.

We may deduce the optimal exponent $\lambda$ for given $\omega$ and
$\alpha$ by linearizing equation (\ref{angeom}) about $\theta=0$:
\beq
\label{m9.1}
\delta\theta_{i+1}-\left[2+\frac{1-\omega}{\alpha}\right]\delta\theta_i
+\delta\theta_{i-1}=0.
\eeq
The general solution to (\ref{m9.1}) is $\delta\theta_i=A\lambda^i+
B\lambda^{-i}$ where
\beq
\label{m9.2}
\lambda=1+\frac{1-\omega}{2\alpha}+\left[\frac{1-\omega}{\alpha}+
\left(\frac{1-\omega}{2\alpha}\right)^2\right]^\frac{1}{2}.
\eeq
It is this type of exponential decay, $|\theta_i|\sim\lambda^{-|i|}$, which 
we will encounter in our numerical analysis. 

Another interesting piece of information can be deduced from equation
(\ref{m8}), namely the tangent vector to the continuation curve in $\ell_2$,
that is, the direction of continuation away from $\alpha=0$. Substituting
(\ref{m4}) and (\ref{m7}) into (\ref{m8}) at $\alpha=0$ yields
\beq
\label{m10}
\left[\left.\frac{d\theta^\alpha}{d\alpha}\right|_{\alpha=0}\right]_i=
\left\{\begin{array}{cc}
0 & |i|>1 \\
\frac{1+\omega}{\sqrt{1-\omega^2}} & |i|=1 \\
\frac{2}{\sqrt{1-\omega^2}} & i=0.
\end{array}\right.
\eeq
So the continuation initially proceeds by pulling the central spin, $\bn_0$, further
away from the vacuum $\vac$, and pulling its nearest neighbours, $\bn_{\pm 1}$ away from
$\vac$ in the same direction (as $\nvec_0$ and each other), while leaving
all other spins fixed at $\evec_3$, to first order.

We have proved the existence of a continuation of the one site breather
$\theta^b$, but the same argument can be applied to any zero of
$F(\cdot,0):\ell_2\ra\ell_2$ to prove existence of more general discrete
breathers at small $\alpha$. Clearly, if $F(\theta,0)=0$ then
$\theta_i\in\pi\Z\cup\cos^{-1}\{\omega\}$ for all $i$, and given that
$\theta\in\ell_2$, $\theta_i=0$ for all $|i|$ sufficiently large. So
one can construct frequency $\omega$ solutions at $\alpha=0$ like
\beq
\label{m11}
\wt{\theta}=(\ldots,0,0,\cos^{-1}\omega,\pi,\pi,\pi,\cos^{-1}\omega,0,0,
\ldots)
\eeq
for example. It is easy to check that $DF_{(\wt{\theta},0)}$ is an 
isomorphism, so an analogous result to the theorem above applies here
also. Of particular relevance to the numerical work in section \ref{sec-numeric}
will be those periodic solutions obtained by continuing $\alpha=0$ solutions
with even reflection symmetry about
$i=0$, that is, $\theta_{-i}\equiv\theta_i$. Such a solution may be
specified by a finite coding sequence \cite{aubseq} representing the
values of $\theta_i$ for $i=0,1,2,\ldots$ by $+$ if $\theta_i=0$,
$o$ if $\theta_i=\cos^{-1}\omega$ and $-$ if $\theta_i=\pi$, with the 
convention that the last symbol represents the constant tail of the
sequence. Of course, since $\theta\in\ell_2$, each coding sequence must end
in $+$, so the last symbol is redundant. Nevertheless, we include it in
order to make the notation more suggestive. The symbols themselves are
meant to represent the directions of the central spins:
up is $+$, down $-$, while $o$ is somewhere ($\omega$
dependent) in between. We shall refer to those breathers obtained by
continuing such a solution by the same coding sequence. Hence 
$\theta^\alpha$, the continuation of $\theta^b$ is a $(o+)$-breather, while
$\wt{\theta}$ and its continuation are $(--o+)$-breathers, for example.

\section{Numerical results on discrete breathers}\news
\label{sec-numeric}
In this section we present the results of an extensive numerical investigation
of the continuation of 1-site and various multi-site breathers from the 
anti-continuum limit. To construct numerical solutions we must first truncate
to a finite number of lattice sites, which we implement by fixing vacuum boundary
conditions, $\theta_i=0$ for $i=\pm N$, and restricting to the interior of the
lattice $|i|<N.$ Furthermore, the solutions we consider are symmetric about the
central lattice site, that is, $\theta_i=\theta_{-i}$, so only the sites $i\ge 0$
need be considered, with an appropriate modification at the central site $i=0$.
Explicitly, the task of finding a numerical solution reduces
to finding a zero of the following $(N+1)$-component vector
\be
\label{f}
F_i=
\left\{
\begin{array}{ll}
\alpha[\cs_i(\sn_{i+1}+\sn_{i-1})-\sn_i(\cs_{i+1}+\cs_{i-1})]+\sn_i(\omega-\cs_i)
& 0<i<N \\
2\alpha[\cs_0\sn_1-\sn_0\cs_1]+\sn_0(\omega-\cs_0) & \ i=0 \\
\theta_N & \ i=N 
\end{array}\right.
\end{equation}
as a function of the $(N+1)$ values $\theta_i$, $i=0,..,N.$

To find a zero of the vector $F_i$, for a given value of $\alpha$, we begin with
the value $\alpha=0$, where we have the explicit solution corresponding to the
 trivial 1-site breather (\ref{m4}) (and later we use other multi-site 
breathers).
We then increase $\alpha$ by a small amount and use a Newton-Raphson scheme 
to converge to the new solution. This process is repeated until the desired
 value of $\alpha$
has been obtained. During the calculation the determinant of the Jacobian matrix
$J_{ij}={\partial F_i}/{\partial \theta_j}$ is monitored to ensure that it is 
non-zero,
as a check that the scheme is not accidentally jumping to a different solution
 branch.
The results presented below were obtained from the value $N=20$, though different size
lattices were checked to confirm that, providing $N$ is sufficiently large, the results
are not sensitive to the number of sites.

\begin{figure}[ht]
\centerline{
\epsfxsize=7cm
\epsfbox[50 50 407 301]{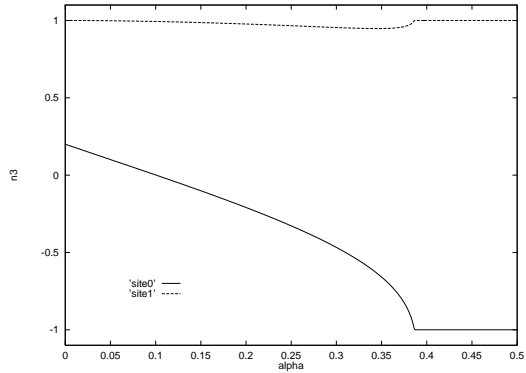}}
\caption{Plots of $\cs_0$ and $\cs_1$ for the continuation of the $(o+)$-breather
with $\omega=0.2$, from $\alpha=0$ to $\alpha=0.5$.}
\label{fig1}
\end{figure}

In Figure 1 we display the results of the continuation of the $(o+)$-breather,
with frequency $\omega=0.2$, from $\alpha=0$ to $\alpha=0.5.$ Presented are
plots of $\cs_0$ and  $\cs_1$, that is, the third component of the spin
for the central and next to central sites. From this plot it is clear that the 
$(o+)$-breather continues until it joins the static $(-+)$ solution. This is true 
for all
frequencies in the band $\omega\in(-1,0.424]$, where we have computed the edge
of the band to within the numerical accuracy given. It is not surprising that
for $\omega$ close to $-1$ the $(o+)$-breather continues to the static $(-+)$
 solution
since for $\alpha=0$ the $(o+)$-breather tends to the $(-+)$ solution as
$\omega\rightarrow -1.$

\begin{figure}[ht]
\centerline{
\epsfxsize=7cm
\epsfbox[50 50 407 301]{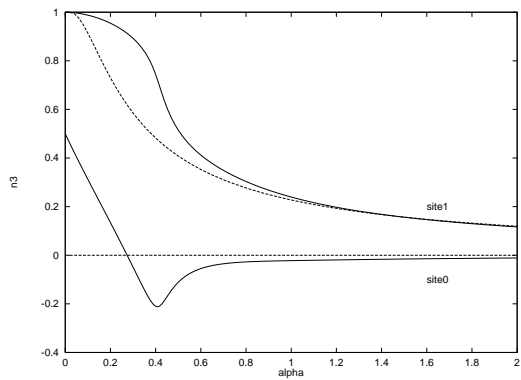}}
\caption{Plots of $\cs_0$ and $\cs_1$ for the continuation of the $(o+)$-breather
(solid curves) with $\omega=0.5$, from $\alpha=0$ to $\alpha=2$.
The dashed curves are the corresponding quantities for the soliton solution.}
\label{fig2}

\end{figure}
\begin{figure}[ht]
\centerline{
\epsfxsize=7cm
\epsfbox[50 50 407 301]{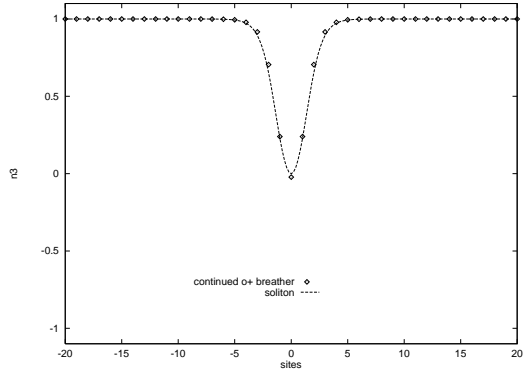}}
\caption{
$\cs_i$ for all the lattice sites of the $\omega=0.5$  $(o+)$-breather
with $\alpha=2$ (diamonds). The dashed curve is the corresponding soliton solution.
}
\label{fig3}

\end{figure}
In Figure 2 we present the results of a similar calculation, but this time the
initial $(o+)$-breather has a frequency $\omega=0.5.$ The values of $\cs_0$ and  
$\cs_1$ are shown as solid curves and the dashed curves represent the same
quantities for the continuum soliton solution (\ref{soliton}) sampled at lattice
sites $x=0$ and $x=h$, where $h=1/\sqrt{\alpha}.$ As the coupling $\alpha$
is increased, the values at these sites tend towards those of the continuum 
soliton solution, demonstrating that the $(o+)$-breather approaches the 
soliton solution in the continuum limit. As further evidence, in Figure 3 we plot
$\cs_i$ for all lattice sites, for the value $\alpha=2$, superimposed with the soliton
solution given in (\ref{soliton}), with $x=i/\sqrt{2}.$ As a final check, starting
with initial conditions obtained from sampling the soliton solution at the lattice
sites and continuing backwards to $\alpha=0$, we find that we recover the $(o+)$-breather.
We have also verified that the decay rate of the $(o+)$-breather shown in
Figure 3 fits extremely well to the derived exponential decay constant given in (\ref{m9.2}).  

Similar results apply for all frequencies in the band $\omega\in(0.424,1)$, 
with the $(o+)$-breather continuing to the soliton solution, although the region in
which $\omega$ is close to 1 is numerically inaccessible within the present scheme 
since the soliton has a relatively slow spatial decay in this region.

Just below the interface of the two bands, that is for $0\le 0.424-\omega\ll 1$,
there is a more complicated bifurcation structure between the $(o+)$-breather
and the static $(-+)$ solution than in the bulk of the lower band $(-1,0.424].$
To examine this region requires a slightly more sophisticated numerical approach,
as we now describe, by illustrating the case $\omega=0.42.$ 

\begin{figure}[ht]
\centerline{
\epsfxsize=7cm
\epsfbox[50 50 407 301]{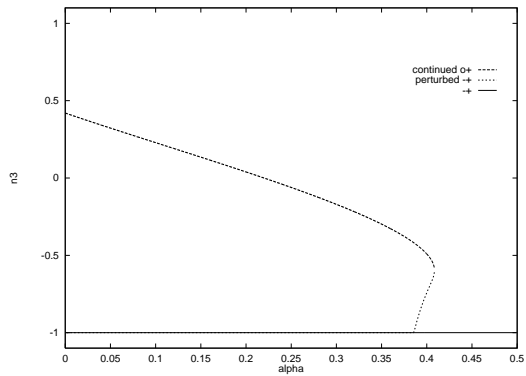}}
\caption{
$\cs_0$ for the continuation of the $(o+)$-breather (dashed curve), the
static $(-+)$-solution (solid line), and the perturbed $(-+)$-solution 
(dotted curve).
}
\label{fig4}

\end{figure}
\begin{figure}[ht]
\centerline{
\epsfxsize=7cm
\epsfbox[150 230 470 550]{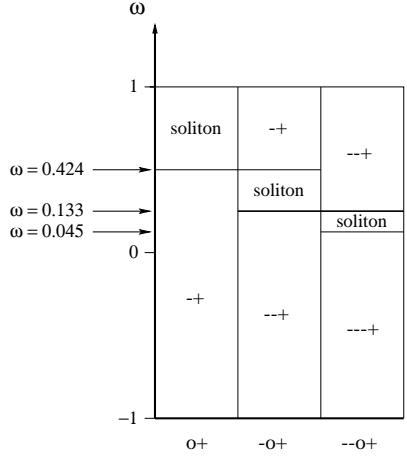}}
\caption{
A table displaying the results, as $\omega$ is varied, of continuing
breathers of type $(o+)$,$(-o+)$ and $(--o+).$ The symbol at the foot
of each column indicates which breather is being continued and the
symbol inside the column represents the end result of the continuation
at that frequency.
}
\label{fig5}

\end{figure}
\begin{figure}[ht]
\centerline{
\epsfxsize=12cm
\epsfbox[25 270 580 510]{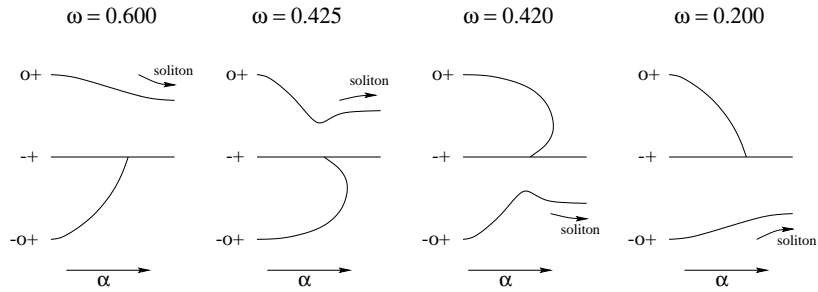}}
\caption{
A schematic to indicate how the bifurcation pattern 
between the $(-+)$ solution and the $(o+)$ and $(-o+)$
breathers varies with $\omega.$
}
\label{fig6}

\end{figure}
For $\omega=0.42$ the continuation of the $(o+)$-breather is presented in Figure 4,
where the dashed line represents the value of $\cs_0.$ The continuation fails at
$\alpha=\alpha^*=0.4085$, where the determinant of the Jacobian matrix $J$ is zero.
We expect that this solution still bifurcates from the static $(-+)$ solution,
but that it first turns around in $\alpha$ space. To confirm this expectation
we need to find the value $\alpha=\widehat\alpha$, at which the bifurcation of the $(-+)$
 solution takes place, and compute the tangent vector in the direction of this
bifurcation. The Jacobian of the $(-+)$ solution is the tridiagonal matrix
\be
J=\pmatrix{ 
2\alpha-\omega-1 & -2\alpha &  &  & &  \cr
-\alpha & \omega-1 & \alpha &   &   &  \cr
 & \alpha & -2\alpha+\omega-1 & \alpha &   &  \cr
 & & \ldots & \ldots & \ldots & \cr
 & & & \alpha & -2\alpha+\omega-1 & \alpha \cr
 & & & &0 & 1\cr
}\ee
Computing the determinant of this matrix (for $N=20$) and setting $\omega=0.42$, we find that
$\mbox{det}\,J$ has only one real and positive root, which occurs at
 $\alpha=\widehat\alpha=0.3858.$ Note that this calculation provides the first check on
our bifurcation assumption since we require that $\widehat\alpha<\alpha^*$, which is indeed
true. Substituting $\alpha=\widehat\alpha$ into $J$ we then compute the eigenvector,
$\delta\theta$, corresponding to the zero eigenvalue, normalized so that its norm
$|\delta\theta|^2=\epsilon\ll 1.$ Finally, we take the $(-+)$ solution, in which $\theta_0=\pi$ and
$\theta_i=0$ for $i>0$, and add the perturbation $\delta\theta$, to create an initial condition
which we then iterate to converge to a solution
for a value of $\alpha$ obtained by incrementing $\widehat\alpha$ by a small amount.
Following this solution for increasing $\alpha$ we find that it continues until $\alpha=\alpha^*$,
where it meets the end point of the continued $(o+)$-breather. This is illustrated in
Figure 4, where we plot $\cs_0$ (dotted curve) for the solution branch obtained by
perturbation of the static $(-+)$ solution, as just described. Following the end point of this
solution branch backwards in $\alpha$ confirms that it joins the static $(-+)$ solution at
$\alpha=\widehat\alpha.$

To summarize, the results of the continuation of the $(o+)$-breather, as $\omega$ is varied,
are represented in the first column of Figure 5, and presented schematically in the upper
half of Figure 6. For $\omega\in(0.424,1)$ the $(o+)$-breather continues to the soliton,
for $\omega$ outside this range, but close to $0.424$, it turns back and joins the
static $(-+)$ solution, and in the bulk of the band $(-1,0.424]$ it joins the
static $(-+)$ solution without turning back.

\begin{figure}[ht]
\centerline{
\epsfxsize=7cm
\epsfbox[50 50 407 301]{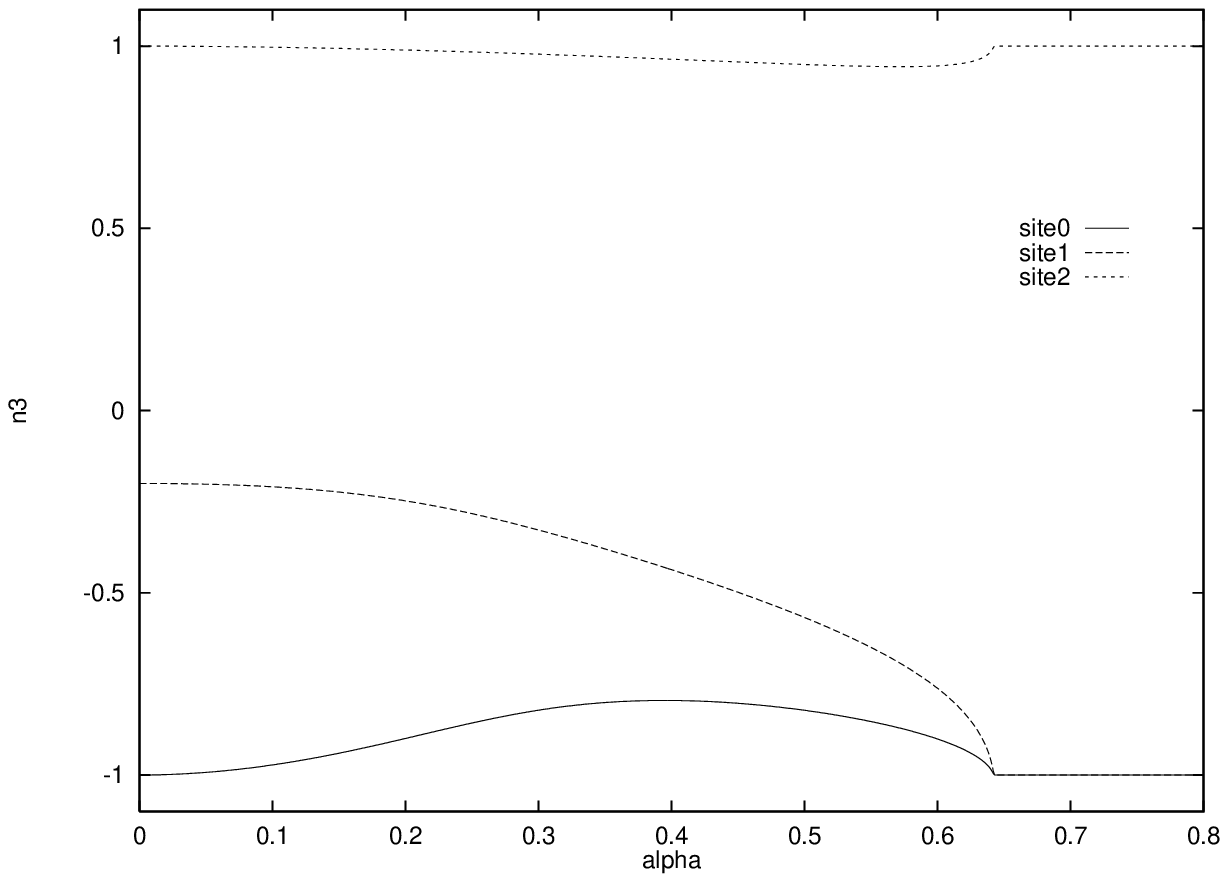}}
\caption{
$\cs_0,\cs_1,\cs_2$ for the continuation of the $\omega=-0.2$ 
$(-o+)$-breather for $\alpha\in[0,0.8].$
}
\label{fig7}

\end{figure}
For $\omega$ above $0.424$ there is no longer a bifurcation of the static $(-+)$
solution with the $(o+)$-breather, but to fill in the full bifurcation pattern
of the $(-+)$ solution we must now consider the continuation of the $(-o+)$-breather.
Note that now the start point of the continuation, $\alpha=0$, is already a multi-site breather, 
since we have applied reflection symmetric conditions, $\theta_i=\theta_{-i}$,
 and the breathing site is not located at $i=0.$

\begin{figure}[ht]
\centerline{
\epsfxsize=7cm
\epsfbox[50 50 407 301]{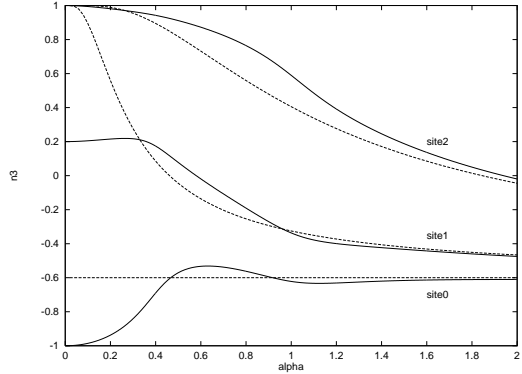}}
\caption{
$\cs_0,\cs_1,\cs_2$  (solid curves) for the continuation of the $\omega=0.2$ 
$(-o+)$-breather for $\alpha\in[0,2].$ 
The dashed curves are the corresponding quantities for the soliton solution.
}
\label{fig8}

\end{figure}
\begin{figure}[ht]
\centerline{
\epsfxsize=7cm
\epsfbox[50 50 407 301]{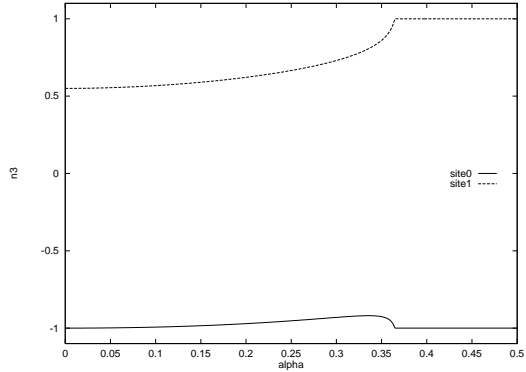}}
\caption{
$\cs_0,\cs_1$ for the continuation of the $\omega=0.55$ 
$(-o+)$-breather for $\alpha\in[0,0.5].$
}
\label{fig9}

\end{figure}
In Figure 7 we plot $\cs_i$ for the first three sites, $i=0,1,2$, for the continuation
of the $(-o+)$-breather with frequency $\omega=-0.2.$ Clearly, for this frequency, the
$(-o+)$-breather continues to the static $(--+)$ solution, and this is true for all
$\omega\in(-1,0.133).$ From a similar earlier discussion, it follows that this result
is to be expected, at least for $\omega$ near $-1.$
For $\omega\in[0.133,0.424]$ the $(-o+)$-breather approaches the soliton
solution in the continuum limit. As an example, for $\omega=0.2$, we present in
Figure 8, a plot of $\cs_i$ for the first three sites (solid curves) and also
the corresponding quantities for the sampled soliton solution (dashed curves).
For $\omega\in(0.424,1)$ the $(-o+)$-breather continues to the static $(-+)$ solution,
as demonstrated in Figure 9 for $\omega=0.55.$ These results are summarized in the
second column of Figure 5.

It is, of course, no accident that the frequency ($\omega=0.424$) at which the $(o+)$-breather
fails to continue to the soliton is precisely that at which the $(-o+)$-breather begins
to continue to the soliton. This can be understood by completing the bifurcation pattern
of the $(-+)$ solution, using the above results on the continuation of the
$(-o+)$-breather, to fill in the bottom half of the schematic in Figure 6.
As presented in the schematic, there is always a bifurcation of the $(-+)$ solution,
but it switches over from the $(-o+)$-breather to the $(o+)$-breather as the
frequency is decreased to the critical value $\omega=0.424.$ At this point the
$(o+)$-breather no longer continues to the soliton, but the $(-o+)$-breather now does.
We have verified this structure with a number of detailed further calculations, for example,
we have confirmed the turning back of the continuation of the $(-o+)$-breather by 
perturbing the $(-+)$ solution as described earlier.

The lower edge of the band $\omega\in[0.133,0.424]$ can also be understood in a similar
fashion by an analysis of the bifurcation pattern of the $(--+)$ solution, which requires
computations of the continuation of the $(--o+)$-breather. These calculations have been
performed and the results are summarized in the third column of Figure 5. 
Again there is a soliton band, which begins at the frequency $\omega=0.045$,
and ends at the start of the $(-o+)$-breather soliton band where
$\omega=0.133.$ Below the soliton band the $(--o+)$-breather continues to the
$(---+)$ solution and above the soliton band it continues to the $(--+)$ solution.
Computations confirm that the bifurcation pattern of the $(--+)$ solution
is analogous to that of the $(-+)$ solution depicted in Figure 6, with an extra
$-$ sign inserted into the coding sequence of each solution together with a shift in the frequencies.

Given the above results we are naturally led to the conjecture that the pattern
of soliton bands continues, with each band covering a smaller range of frequencies,
but such that for any frequency $\omega\in(0,1)$ there is a discrete breather solution
whose continuation from $\alpha=0$ tends to the soliton solution
in the continuum limit $\alpha\rightarrow\infty.$

Generally, as mentioned in the introduction, discrete breathers in networks of coupled oscillators
fail to continue beyond a certain coupling due to resonance with phonons. In the system 
considered here, the role of phonons is played by spin waves,
\beq
\delta\nvec_i(t)=\left(\begin{array}{c}
\cos(ki-\omega_p t) \\ \sin(ki-\omega_p t) \\ 0
\end{array}\right),
\eeq
which are travelling wave solutions of (\ref{eom2}) linearized about the vacuum
$\nvec=\evec_3$. These have dispersion relation
\be
\omega_p=1+4\alpha\sin^2(k/2),
\ee
so the spin waves form a frequency band with $\omega_p\in[1,1+4\alpha].$
Note that the maximal frequency spin wave ($k=\pi$) is standing, hence of the form
(\ref{ansatz}) with frequency $\omega_p=1+4\alpha$. It  may equally well be regarded as 
periodic of frequency $\omega_p/n$ for any $n\in\Z^+$. Mathematically, a bifurcation in
the continuation of breathers may only occur where $DF$ acquires nontrivial kernel, that is
the equation of ``motion'' (\ref{angeom}) linearized about the breather supports a nonzero
solution. Of course phonons, being linearized solutions about the {\em vacuum}, 
never lie in
${\rm ker}\, DF$ strictly speaking. 
Nevertheless, since the breather approaches the vacuum exponentially fast as 
$|i|\ra\infty$, it is generally accepted that the existence of a standing phonon of 
frequency $\omega_p=n\omega$ generically implies nontriviality of ${\rm ker}\, DF$, that 
is, the phonon is close to a tangent vector in ${\rm ker}\, DF$, approaching it 
asymptotically as $|i|\ra\infty$. In the present case, the standing spin wave lies outside
$\ell_2$, so cannot be close to ${\rm ker}\, DF$, and hence cannot cause a bifurcation.
In practice this technical point is irrelevant: the numerics are performed on
a finite chain where such distinctions are impossible to make. Why then do the standing
spin waves cause no bifurcations in the numerical continuation? The reason is that our
map $F$ contains explicit parametric dependence on $\omega$, so only the standing spin
wave of frequency $\omega_p=\omega$ is in the kernel of $DF$ at the vacuum. Higher 
harmonics, $\omega_p=n\omega$ are {\em not} linearized 
solutions of our
ansatz derived equation (\ref{angeom}), and so cannot cause bifurcations.
This is entirely consistent with physical intuition. These breathers, like those
in the DNLS system, are monochromatic: their time dependence contains no higher harmonics so one 
does not expect them to resonate with the spin waves. It is precisely the imposition of
a monochromatic ansatz (\ref{ansatz}) which introduced the parametric $\omega$ dependence 
discussed above.

\section{Linear stability}\news
\label{sec-stability}

Linear stability of breathers in networks of anharmonic oscillators was studied by
Aubry in \cite{aubstab}. Despite the many superficial differences between spin chains and
oscillator networks, we will find that the analytic framework developed by Aubry readily 
adapts to this new setting, yielding similar results. In particular, we will prove that
$(o+)$-breathers of all frequencies $\omega\in(-1,0)\cup(0,1)$
must be linearly stable for sufficiently small $\alpha$ provided $\omega^{-1}\notin\Z$.
Just how small $\alpha$ must be, and the stability properties of the various other
breather types will be investigated numerically.

As in \cite{aubstab}, it is technically convenient to truncate the lattice to (large but)
finite size $N$ (the results will be independent of $N$). 
Existence of discrete breathers in such a system follows from an identical argument to that of
theorem \ref{exist}, but with $\ell_2$ replaced by $\R^N$ with its usual norm.
Equation (\ref{eom2}) along
with fixed endpoints $\nvec_0=\nvec_{N+1}=0$ then define a flow on phase space
$M=(S^2)^N$. Crucial to the analysis is the fact that this flow is Hamiltonian with
respect to the natural symplectic structure on $M$, namely
\beq
\label{sympform}
\Omega=\sum_i\Omega_i
\eeq
where $\Omega_i$ is the area form on the $i$-th 2-sphere. The Hamiltonian 
$H:M\ra\R$ is the truncation of (\ref{ham}), again with fixed endpoints. It follows that
the flow is symplectomorphic.

A period $T$ solution of the system is a fixed point of the return map $P_T:M\ra M$,
$P_T:n(0)\mapsto n(T)$. Such a solution is said to be linearly stable if the
spectrum of its associated
Floquet map $\fl=(dP_T)_{n(0)}:T_{n(0)}M\ra T_{n(T)=n(0)}M$ lies within the closed unit
disk $D=\{z\in\C:|z|\leq 1\}$. Recall that $\fl:\delta n(0)\mapsto\delta n(T)$ where $\delta n(t)$ is
the solution of the linearization of (\ref{eom2}) about the solution $n(t)$, explicitly,
\beq
\label{leom}
\delta\dot{\nvec}_i=\alpha[\nvec_i\times(\delta\nvec_{i+1}+\delta\nvec_{i-1})+\delta\nvec_i
\times(\nvec_{i+1}+\nvec_{i-1})]+(\delta\nvec_i\cdot\evec_3)\nvec_i\times\evec_3.
\eeq
Since $P_T$ is a symplectomorphism, $\fl$ is a symplectic map, that is
\beq
\Omega(\fl\, \delta n,\fl\, \delta n')\equiv\Omega(\delta n,\delta n').
\eeq
It follows that if $\lambda\in\spec\fl$, so are $\bar{\lambda}$, $1/\lambda$ and
$1/\bar{\lambda}$. Hence the solution is linearly stable if and only if $\spec\fl$ lies
on the unit circle $\cd D=\{z\in\C:|z|=1\}$. 

It is straightforward to construct $\fl$ explicitly for all the breathers considered
in sections \ref{sec-analytic} and \ref{sec-numeric}
 in the uncoupled limit, $\alpha=0$. For each $\nvec_i(0)
=(\sin\theta_i,0,\cos\theta_i)$ define the ordered orthonormal basis
\beq
\label{basis}
(\epsvec_i=(\cos\theta_i,0,\sin\theta_i), \evec_2)
\eeq
for $T_{\nvec_i(0)}S^2$, so that 
$T_{n(0)}M=\bigoplus_{i}\spn\langle\epsvec_i,\evec_2\rangle.$ Relative to this basis,
the symplectic form has the usual block matrix expression, namely
\beq
\Omega=\diag(\ldots,R(\frac{\pi}{2}), R(\frac{\pi}{2}),\ldots)
\eeq
where $R(\psi)$ is the $SO(2)$ matrix which performs a clockwise rotation through angle
$\psi$, explicitly,
\beq
\label{Rdef}
R(\psi)=\left(\begin{array}{cc}
\cos \psi & \sin \psi \\
-\sin \psi & \cos \psi \end{array}\right).
\eeq
The
Floquet matrix of the frequency $\omega$ $(o+)$-breather is
\beq
\label{flo+}
\fl=\diag(\ldots,R(T),R(T),S,R(T),R(T),\ldots)
\eeq
where $T=2\pi/\omega$ and $S$ is the matrix
\beq
S=\left(\begin{array}{cc}
1 & 0 \\ (1-\omega^2)T & 1 \end{array}\right). 
\eeq
Similarly, the $(-o+)$-breather has Floquet matrix
\beq
\label{fl-o+}
\fl=\diag(\ldots,R(T),R(T),S,R(T),S,R(T),R(T),\ldots)
\eeq
and so on. Clearly all these matrices have $\spec\fl=\{e^{\pm iT},1\}\subset\cd D$ so the
uncoupled breathers of all types are linearly stable. 

As $\alpha$ increases from 0, theorem \ref{exist} guarantees the existence of a continuous
family of symplectic maps $\fl_\alpha$ (recall that $\alpha\mapsto\theta_\alpha$ is
$C^1$) whose eigenvalues depend continuously on $\alpha$. Do these eigenvalues remain on 
$\cd D$? Clearly only coincident pairs may leave $\cd D$ in tandem, by the symmetry
properties of $\spec\fl$. Krein theory \cite{aubstab} states that an eigenvalue
$\lambda\in \cd D$ may bifurcate off the unit circle only if its associated Krein 
signature is indefinite. This signature is defined as follows. Let $E_\lambda\subset
\C^{2N}$ be the $\lambda$ eigenspace of $\fl$ and $V_\lambda:=\re(E_\lambda\oplus 
E_{1/\lambda})\subset\R^{2N}$. Then on $V_\lambda$ one defines a bilinear form
$Q_\lambda:V_\lambda\oplus V_\lambda\ra\R$ such that
\beq
Q_\lambda(\xi,\eta):=\Omega(\xi,\fl\, \eta).
\eeq
The Krein signature of $\lambda$ is the signature of this bilinear form. Hence, $\lambda$
may bifurcate off $\cd D$ only if $Q_\lambda$ is indefinite, that is, the diagonal map
$\xi\mapsto Q_\lambda(\xi,\xi)$ is neither positive definite nor negative definite on
$V_\lambda\backslash\{0\}$. 

Applying this theory to (\ref{flo+}) one sees that $V_{e^{iT}}\equiv V_{e^{-iT}}\cong
\R^{2N-2}$, $V_1\cong\R^2$, 
\beq
Q_{e^{iT}}(\xi,\xi)=-(\sin T)\, \xi^{T}\xi,\quad
Q_1(\xi,\xi)=(1-\omega^2)\xi_1^2.
\eeq
Provided $\sin T\neq 0$, i.e. $\omega^{-1}\notin\Z$, $Q_{e^{iT}}$ is definite and hence
all the eigenvalues $e^{\pm iT}$ must remain on $\cd D$, at least for sufficiently small
$\alpha$. On the other hand, $Q_1$ is only positive {\em semi}-definite, so no such
constraint applies to the double eigenvalue $\lambda=1$. Note however that for any
nonstatic period $T$ solution $n(t)$ of an autonomous dynamical system, $\dot{n}(0)$ is
an eigenvector of $\fl$ with $\fl\, \dot{n}(0)=\dot{n}(0)$ by time translation invariance.
It follows that one copy of the eigenvalue $\lambda=1$ is fixed for all $\alpha$. The
second is also fixed for sufficiently small $\alpha$ since to move it would have to
(at least) double up due to the symmetry properties of $\spec\fl$. Hence, we have proved:

\begin{thm} \label{stable}
For each $N\in\Z^+$ and 
for all $\omega\in (-1,1)\backslash\{0\}$ such that
$\omega^{-1}\notin\Z$, there exists $\wt{\alpha}(\omega)>0$ such that for all $\alpha\in
[0,\wt{\alpha}(\omega))$ the continued $(o+)$-breathers, whose existence on the $N$ site 
lattice is 
guaranteed by theorem \ref{exist}, are linearly stable.
\end{thm}

Unfortunately, this argument cannot be applied to multisite breathers such as $(-o+)$.
This is because $V_1$ is now (at least) 4 dimensional and $Q_1$ remains only
semidefinite. Time translation symmetry is not sufficient to fix all 4 copies of $\lambda
=1$ under perturbation of $\alpha$, so while the $\alpha=0$ $(-o+)$-, $(--o+)$- etc.\, 
breathers are all stable, instability may develop for arbitrarily small $\alpha$. Note,
however, that the Krein criterion is necessary but {\em not sufficient}: it does not
guarantee instability of $\alpha>0$ multisite breathers, nor of $(o+)$-breathers with
$\omega^{-1}\in\Z$. To investigate these issues, we must resort once again to numerical
analysis.

One may vastly simplify the task of numerically constructing the Floquet matrix
$\fl$ for this problem by transforming to a co-rotating coordinate system. Let
$\rot:\R\ra SO(3)$ such that
\beq
\rot(t)=\diag(R(\omega t),1)
\eeq
where $R(t)$ is the the $SO(2)$ matrix defined in equation (\ref{Rdef}), and
define new variables $\uvec_i(t)$ such that
\beq
\nvec_i(t)=\rot(t)\uvec_i(t).
\eeq
Then the equations of motion become
\beq
\label{eom3}
\dot{\uvec_i}=\alpha\uvec_i\times(\uvec_{i+1}+\uvec_{i-1})+(\uvec_i\cdot\vac)(\uvec_i\times\vac)
-{\cal A}\uvec_i
\eeq
where ${\cal A}=(\rot^{-1}\dot{\rot})(t)\equiv(\rot^{-1}\dot{\rot})(0)\in
so(3)$ is a constant antisymmetric matrix. 
The point is that all the breather solutions we have been considering 
are {\em static} in this coordinate system, namely $\uvec_i(t)=(\sin\theta_i,
0,\cos\theta_i)$, where $\theta_i$ satisfies (\ref{angeom}) as before.
Consequently, the linearized flow is defined by an {\em autonomous} linear
system of ODEs,
\beq
\delta\dot{\uvec}_i=\alpha[\uvec_i\times(\delta\uvec_{i+1}+\delta\uvec_{i-1})+\delta\uvec_i
\times(\uvec_{i+1}+\uvec_{i-1})]+(\delta\uvec_i\cdot\evec_3)\uvec_i\times\evec_3
-{\cal A}\delta\uvec_i.
\eeq
Using the basis $\{\epsvec_{i},\evec_2\}$ for $T_{\uvec_i}S^2$ and writing
\beq
\delta\uvec_i=a_i(t)\epsvec_i+b_i(t)\evec_2,
\eeq
this system reduces to
\beq
\left(\begin{array}{c}\dot{a} \\ \dot{b}\end{array}\right)=
\Lambda\left(\begin{array}{c} a \\ b \end{array}\right)=
\left(\begin{array}{cc} 0 & \Lambda^+ \\ \Lambda^- & 0 \end{array}\right)
\left(\begin{array}{c} a \\ b \end{array}\right)
\eeq
where $\Lambda^\pm$ are constant, tridiagonal, 
symmetric real $N\times N$ matrices with components
\bea
\Lambda^+_{ij}(\theta)&=& \alpha\{-\delta_{i,j+1}-\delta_{i,j-1}+[\cos(\theta_i-\theta_{i+1})
+\cos(\theta_i-\theta_{i-1})]\delta_{i,j}\} \nonumber \\
& & \qquad+\cos\theta_i(\cos\theta_i-\omega)\delta_{i,j},
 \nonumber \\
\Lambda^-_{ij}(\theta)&=& \alpha\{\cos(\theta_i-\theta_j)(\delta_{i,j+1}+\delta_{i,j-1})
-[\cos(\theta_i-\theta_{i+1})
+\cos(\theta_i-\theta_{i-1})]\delta_{i,j}\}
\nonumber \\
\label{***}
& &\qquad + (\omega\cos\theta_i-\cos 2\theta_i)\delta_{i,j}.
\eea
The Floquet matrix $\fl:\delta\nvec(0)\mapsto\delta\nvec(T)$ is simply
\beq
\label{**}
\fl=\rot(T)\exp(T\Lambda)\rot(0)^{-1}\equiv\exp(T\Lambda).
\eeq
One may easily check that equations (\ref{***}), (\ref{**}) reproduce the
$\alpha=0$ matrices previously obtained. Note that we have {\em not} imposed a monochromatic
 ansatz for the
perturbation: $\delta\uvec_i$ and hence $\delta\nvec_i$ are permitted to have arbitrary
time dependence. We have merely made a particularly convenient choice of coordinates.

So a period $T$ breather is linearly stable if and only if its corresponding
$\Lambda$ matrix has $\spec\Lambda\subset i\R\subset\C$, a criterion which is 
trivial to test numerically.
This should be contrasted with the procedure
usually employed in numerical linear stability analyses, where construction of $\fl$
requires a coupled system of $2N(N+1)$ linear and nonlinear ODEs to be 
solved numerically. Of course, the transformation to co-rotating coordinates, and the consequent
 simplification of the Floquet problem, are only possible because we are considering 
anisotropic spin
chains which retain a $SO(2)$ isotropy group. The technique will not work
in chains with fully anisotropic exchange interaction, as are considered in
\cite{flazol}.

The numerical results described below were obtained by computing 
$\spec\Lambda$
using various standard routines taken from \cite{numrec} on a 41 site 
lattice (the results do not differ significantly from those obtained with an
81 site lattice). It should be noted that, while the algorithm to construct the breather
initial profile $\theta$ employed spatial reflexion symmetry, no such symmetry
is imposed on the spectral problem for $\Lambda$. 
Hence {\em all} possible
modes of instability are included in the analysis. Given the spectrum $\{\lambda_i\}$ of 
$\Lambda$, one constructs
\beq
\label{mudef}
\mu:=\max_i(\re\lambda_i)^2.
\eeq
The corresponding breather is stable if and only if $\mu=0$.

\begin{figure}[ht]
\centerline{
\epsfxsize=7cm
\epsfbox[50 180 570 620]{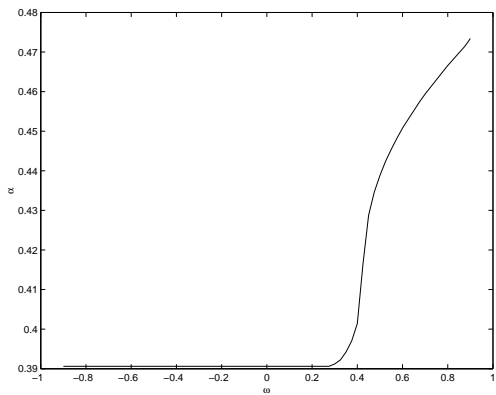}}
\caption{The coupling of first instability $\wt{\alpha}$ as a function of frequency, $\omega$
for $(o+)$-breathers.}
\label{fig10}

\end{figure}

Figure \ref{fig10}
 presents a graph of the maximal $\wt{\alpha}(\omega)$ of theorem \ref{stable}, 
that is, the coupling at which instability of the
frequency $\omega$ $(o+)$-breather first occurs. Three features should be noted. The first is 
that, contrary to expectation, weakly coupled breathers remain stable even at resonant 
frequencies, $\omega^{-1}\in\Z$. In fact, there seems to be nothing special
about these frequencies at all from the standpoint either of breather existence (see sections
\ref{sec-analytic} and \ref{sec-numeric}) or of stability. Second, $\wt{\alpha}(\omega)$ appears
to be finite for all $\omega$: even when $\omega>0.424$ so that the $(o+)$-breather continues 
all the way to the
continuum soliton (a linearly stable solution of the continuum system), it loses stability
along the way. Third $\wt{\alpha}(\omega)$ varies very little with $\omega$ except for $\omega$
close to 1. It appears to grow unbounded as $\omega\ra 1$, which seems reasonable since the 
$\omega=1$ $(o+)$-``breather'' is nothing but the trivial vacuum $\nvec_i=\evec_3$ which is
stable for all $\alpha$. This limit is numerically inaccessible, however (the breathers tend
to spread out over the whole lattice as $\omega$ grows so that finite size effects dominate), so
one should treat this observation with caution.

\begin{figure}[ht]
\centerline{
\epsfxsize=7cm
\epsfbox[50 180 550 620]{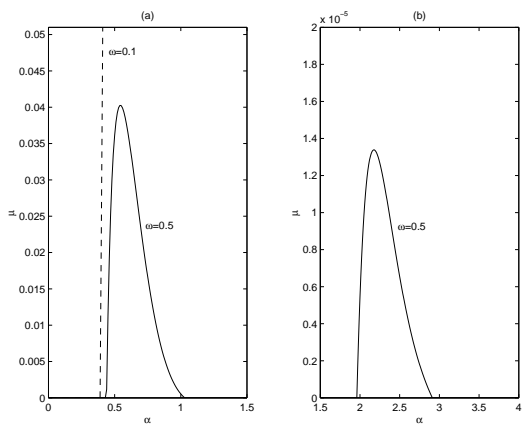}}
\caption{The instability function $\mu(\alpha)$, as defined in equation (\ref{mudef})
 for various $(o+)$-breathers (solid: $\omega=0.5$; dashed $\omega=0.1$). The breather is
stable if and only if $\mu=0$. Note that both are stable for small $\alpha$ despite having 
resonant frequencies.}
\label{fig11}

\end{figure}

Figure \ref{fig11} 
shows plots of $\mu$ against coupling $\alpha$ for $(o+)$-breathers of various
frequencies. For $\omega \ll 0.424$, the breather remains stable until after it has joined the
$(-+)$ branch, only losing stability when this trivial static solution does. After this, $\mu$
grows monotonically with $\alpha$. For $\omega>0.424$
the breather first loses and then regains stability: a hump of compact support
appears in the graph
of $\mu$. This hump is followed by another of roughly the same width but much less tall (note 
the change in vertical scale), so that stability is once again lost and regained. This leads us
to conjecture that instability occurs in regularly repeating bands, the
degree of instability (the height of the humps) decaying exponentially as $\alpha$ increases
and the breather approaches the continuum soliton. It is interesting to note that the transition
from stability to instability (or vice versa) generically coincides with a sign change in the
determinant of $D\hat{F}$, where $\hat{F}$ is the {\em unsymmetrized}
finite lattice  version of the
continuation function $F$ defined in (\ref{m5}). However no such sign change occurs
in the Jacobian of
the {\em symmetrized} continuation function (\ref{f}), as employed in our continuation scheme, 
so the corresponding eigenvector
$\delta\theta\in\ker D\hat{F}$ must have odd spatial parity. It seems likely, therefore, that
stability transitions generically accompany symmetry breaking bifurcations in the full
(i.e.\ unsymmetrized) continuation problem. 

\begin{figure}[ht]
\centerline{
\epsfxsize=7cm
\epsfbox[50 180 550 620]{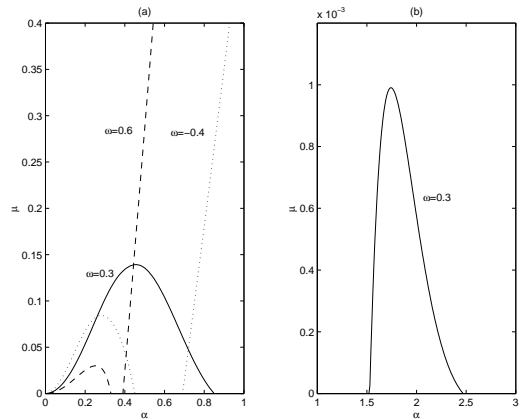}}
\caption{The instability function $\mu(\alpha)$ for various $(-o+)$-breathers
(solid: $\omega=0.3$; dashed: $\omega=0.6$; dotted: $\omega=-0.4$). All are unstable for
small $\alpha$, though they have nonresonant frequency.}
\label{fig12}

\end{figure}

Figure \ref{fig12}
 shows similar plots for $(-o+)$-breathers. Recall that no result equivalent to
theorem \ref{stable} holds in this context, so there is no reason to expect stability even for
arbitrarily small $\alpha$. Indeed, we found that for all $\omega$, $\mu$ departs from $0$
as soon as $\alpha$ does. For $\omega\notin(0.133,0.424)$ the breather remains unstable until
it joins the appropriate trivial branch ($(--+)$ for $\omega<0.133$, $(-+)$ for $\omega>0.424$),
whereupon it becomes briefly stable before instability irrevocably sets in. For $\omega\in
(0.133,0.424)$, the breathers gain stability, then lose it, rather in the fashion of the
$(o+)$-breathers described above. Again, a pattern of repeating bands of diminishing instability
seems to occur as $\alpha$ grows large and the breather approaches the continuum soliton. 
Plots of $\mu(\alpha)$ for $(--o+)$ breathers with $\omega\in(0.045,0.133)$ are very similar:
the breathers start unstable, then gain and lose stability in a repetitive pattern. This leads 
us to conjecture that wherever breathers tend to the continuum soliton as $\alpha\ra\infty$,
a similar banded stability pattern occurs.

\section{Conclusion}\news
In this paper we have proved the existence of discrete breathers in ferromagnetic
spin chains with easy-axis anisotropy, and constructed such breathers by means of
numerical analysis. The main novelty of our numerical results is that discrete
breathers exist independent of the inter-spin coupling $\alpha$, right up to the
continuum limit $\alpha\ra\infty$. This should be contrasted with previous studies
of spin chains with easy-plane anisotropy \cite{WMB}, where the breathers are again 
``monochromatic,'' but the spin waves have a
different dispersion relation, and apparently do cause trouble at large $\alpha$.

The method of proof itself was something of a hybrid: an ansatz was made to reduce
the equation for breathers from a system of ODEs to a purely algebraic system,
but then existence of solutions of the latter was proved by continuing a
decoupled solution in the sequence space $\ell_2$. Consequently, the result shares
nice properties of both the Flach and MacKay-Aubry approaches. Since the reduced
system depends parametrically on $\omega$, spin wave resonances cause no problems,
theoretical or numerical (c.f.\ \cite{aubmac}). On the other hand, the result
generalizes immediately to multi-dimensional spin lattices, as MacKay-Aubry style
theorems usually do. Whether our numerical results generalize similarly, that is, 
whether a picture similar to Figure \ref{fig5} emerges for 2 and 3 dimensional lattices is
an interesting open question. It should be noted that the continuum theory in higher dimensions
does support radially symmetric, exponentially localized soliton solutions of the form 
analogous to (\ref{ansatzc}), namely
\beq
\bn(\xvec,t)=(\sin\theta(|\xvec|)\cos\omega t,-\sin\theta(|\xvec|)\sin\omega t,
\cos\theta(|\xvec|)),
\eeq
although explicit expressions for $\theta(r)$ are not known \cite{KIK}. 

The ansatz may be interpreted as transforming to co-rotating coordinates, wherein the breather
solutions are static. This viewpoint greatly simplifies the linear stability analysis since
one need only solve an autonomous linear system of ODEs. In this way we have made an extensive
study of the linear stability properties of the breathers. The results suggest that those
breather families which converge as $\alpha\ra\infty$ to the continuum soliton experience
repeating bands of instability of diminishing strength as $\alpha$ increases. We found 
that the spin waves have no influence whatsoever on stability issues, even though 
the stability analysis includes all possible perturbations, including those {\em outside} the
monochromatic ansatz. Flach et al also perform a numerical linear stability analysis for 
breathers in their
spin chains, and obtain results broadly similar to ours \cite{flazol}, namely
1-site breathers are found to be stable at weak coupling, and certain multisite breathers are
found to be unstable at weak coupling. The subsequent loss and regain of (in)stability for
increasing coupling is not reported - perhaps it is special to chains with isotropic exchange
interaction. Their survey of the breather existence domains is rather less extensive than ours,
however, presumably due to the much higher computational cost in their systems, so it is
possible that similar band structures to those found here exist there also. 

One open question immediately arises: that of breather mobility in this system. 
In the continuum limit, there exist travelling solitons which
propagate at constant speed \cite{KIK}. 
Can one find analogous propagating breathers at finite coupling? There seems to be
little hope of proving anything rigorously about such objects, but one could still
hope to study breather mobility numerically, focusing on how breather
mobility depends on $\omega$ and $\alpha$. For a survey of what is known 
about moving discrete breathers, see the review articles \cite{flarev,laisie}. One 
striking observation by Aubry and Cretegny is that breather mobility may be associated
with certain behaviour of the Floquet matrix \cite{aubcre}. 
In a one-parameter family of breathers (such as
$\theta^\alpha$), mobility occurs precisely at a value, $\alpha'$ say, where a certain type of 
Krein bifurcation occurs:
$\fl$ becomes defective and  so-called ``marginal modes'' appear (they span the orthogonal
complement of the sum of the $\fl_{\alpha'}$ eigenspaces). 
Perturbing $\theta_{\alpha'}$ in the
direction of such a marginal mode produces a slowly translating breather. So breather mobility 
seems to be naturally associated with transitions from linear stability to instability. It
would be interesting to see whether the stability transitions observed in section
\ref{sec-stability} have such significance.

\section*{Acknowledgements}\news
We acknowledge the EPSRC for a Postdoctoral Research Fellowship (JMS) and an
 Advanced Fellowship (PMS). One of us (JMS) would like to thank Chris Eilbeck for valuable
conversations.\\

\end{document}